\documentclass[aps,twocolumn,pra,showpacs,eqsecnum]{revtex4}
 \usepackage{amsmath,amssymb,latexsym,mathrsfs}
 \usepackage{feynmf}
 \input xypic
 \usepackage[dvips]{graphicx}
 
\begin{document}
\title{
Pseudo-Path Semiclassical Approximation to Transport through Open Quantum 
Billiards: Dyson Equation for Diffractive Scattering}

\author{Christoph Stampfer} 
\altaffiliation[Present address: ]{Chair of Micro and Nanosystems, Swiss Federal Institute of Technology Zurich (ETH Zurich),
 Tannenstr. 3, 8092 Zurich, Switzerland.}
\affiliation{Institute for Theoretical Physics, Vienna University of
Technology, Wiedner Hauptstra\ss e 8-10/136, A-1040 Vienna, Austria, EU}
\author{Ludger Wirtz} 
\affiliation{Institute for Electronics, Microelectronics and Nanotechnology, B.P. 60069, 59692 Villeneuve d'Ascq Cedex, France, EU}

\author{Stefan Rotter}
\affiliation{Institute for Theoretical Physics, Vienna University of
Technology, Wiedner Hauptstra\ss e 8-10/136, A-1040 Vienna, Austria, EU}

\author{Joachim Burgd\"orfer}
\affiliation{Institute for Theoretical Physics, Vienna University of
Technology, Wiedner Hauptstra\ss e 8-10/136, A-1040 Vienna, Austria, EU}
 
\date{\today}
 
\begin{abstract}
We present a semiclassical theory for transport through open billiards of
arbitrary convex shape that includes 
diffractively scattered paths at the lead openings. Starting from a Dyson
equation for the semiclassical Green's function we develop a diagrammatic
expansion that allows a systematic summation over classical paths and pseudo-paths which consist of classical paths joined by diffractive scatterings
(``kinks''). This renders the inclusion of an
exponentially proliferating number of pseudo-path combinations numerically tractable for both regular and chaotic billiards. For a circular billiard and the Bunimovich stadium the path sum
leads to a good agreement with the quantum path length power spectrum up 
to long path length. Furthermore, we find excellent numerical agreement 
with experimental studies of  quantum scattering in microwave billiards
where pseudo-paths provide a significant contribution.\\

\end{abstract}

\maketitle
 
\section{Introduction}
Semiclassical approximations are among the most useful tools in describing and analyzing ballistic transport in mesoscopic systems. On a fundamental level, semiclassical techniques allow to build a bridge between classical and quantum mechanics: the classical paths carry an amplitude which reflects the geometric stability of the orbits and a phase that contains the classical action and accounts for quantum interference \cite{gutz91a,berry72,feyn65}.\\
Ballistic transport through billiards has been studied extensively in the last 
decade \cite{lin93,bog00,yan94,bar93,bar93a,ish95,schw96,vat97,wir97,wir99,blom01,blom02,
blom03,wir03, Aki97,per00,naz01,naz02} and 
a variety of semiclassical approximations 
\cite{bog00,yan94,bar93,schw96,wir97,wir99,blom01,blom02,wir03,arg95,ric94} 
have been introduced in order to provide a qualitative and, in part, also a 
quantitative description of these systems. 
In particular, universal conductance fluctuations (UCF) and the
``weak localization'' (WL) have been studied \cite{bar93} in order to delineate
characteristic differences in the quantum transport of classically
chaotic and integrable billiards. Very recently, quantum shot noise
\cite{blant} in ballistic cavities that are either chaotic \cite{ober,bee}, 
regular \cite{aig} or display a mixed phase space \cite{sim} has been used
as a probe of the quantum-to-classical and chaotic-to regular cross-over
\cite{ober,bee,sim,aig,sil}.

The approach of the (semi) classical limit of ballistic quantum transport is
both conceptually as well as numerically non-trivial as it represents,
generically, a multi-scale problem. The two-dimensional quantum billiard (or
quantum dot, see Fig.~1) is characterized by an area $A$ or linear dimension
$D=\sqrt{A}$. The quantum wires (or leads) to which the billiard is attached
have the width $d$. In order to reach sufficiently long dwell times such that
differences between transiently regular and chaotic motion become important,
the relation $d/D \ll 1$ should hold. To approach the semiclassical limit for
the motion inside the billiard requires $\lambda_D \ll D$ ($\lambda_D$ de
Broglie wave length) or equivalently $kD \gg 2 \pi$. Furthermore, if the
(disorder) potential inside the dot varies over a length scale  $a_P$, we
should require $\lambda_D \ll a_P$ for a semiclassical approximation to hold. 
These conditions pertaining to the dot are
necessary but not sufficient. Since the scattering ($S$) matrix maps
asymptotic scattering states onto each other, also the entrance and exit
channel states in the quantum wire should reach their classical limit,
$\lambda_D \ll d$ or $kd \gg 2 \pi$. The latter limit is virtually impossible to
reach, neither experimentally for quantum dots \cite{mar,45,ober} or microwave
billiards \cite{kim02} nor numerically \cite{rott00}. We will therefore focus
in the following on the ``interior'' or ``intermediate'' semiclassical regime
pertaining to the interior of the billiard, $\lambda_D \ll D$ with convex
hard-walled boundaries such that quantum diffraction in the interior can be
neglected, with the understanding, however, that quantum effects due to the
coupling to the asymptotic quantum wires have to be taken into
account. Accordingly, the term ``semiclassical approximations'' refers in the
following to approximations to the constant-energy Green's function for
propagation in the interior of the dot $G(\vec{r}, \vec{r}', k)$ by an
approximate semiclassical limit, $G^{SC}$ to be discussed below. The
projections of $G$ onto asymptotic scattering states with transverse quantum
numbers $m_L(m_R)$ yield the amplitudes for transmission $T_{m_L, m_R}$ from
the entrance (left) to the exit (right) lead and reflections $R_{m_L,
  m_L'}$. Standard semiclassical approximations to $G$ face several
fundamental difficulties \cite{bar93,wir97,wir99,vat97,ric94,stone94}:
among many others, unitarity is violated with discrepancies in some cases as
large as the conductance fluctuations the theory attempts to describe
\cite{wir99,blom01}. Likewise, the anti-correlation $\delta|T|^2=-\delta|R|^2$
between transmission fluctuations, $\delta|T|^2$, and the
corresponding fluctuations in the reflection, $\delta|R|^2$,
as a function of the wave number $k$ is broken. Also, the ``weak
localization'' effect is considerably overestimated
\cite{bar93a,stone94}. These difficulties are due to the fact that
hard-walled billiards possess ``sharp edges'' at the entrance and exit leads
even though
the interior of the dot features a smooth (in the present case, a constant)
potential. At these sharp edges the contacts to the quantum wires feature
spatial variations of
the potential where the length scale $a_P$ approaches zero. Consequently the
semiclassical limit $\lambda_D/a_P \ll 1$ cannot be reached, no matter how
small $\lambda_D$ (or large $k$) is. In other words, the quantum properties of
the leads influence also the semiclassical dynamics in the interior. This
observation is the starting point for diffractive corrections such as the
Kirchhoff diffraction \cite{schw96} or Fraunhofer diffraction \cite{wir97}.

We have recently developed a pseudo-path semiclassical approximation (PSCA)
\cite{wir03} with pseudo-paths that result from spawning of classical paths
due to
diffractive (i.e.~non-geometric) reflections in the lead mouths (or point
contacts). Pseudo-paths play an essential role when incorporating
indeterministic features into the semiclassical description of transport. 
Their existence has
also recently been pointed out in Ref.~\cite{prange}, although no explicit
numerical investigations were performed.
While classical
trajectories are either ejected through the exit lead contributing to $T$ or
return back to the entrance lead contributing to $R$, a quantum wavepacket
will do both. Pseudo-paths interconnect otherwise disjunct subsets of
classical paths that exit either through the left or right lead. The lack of
this coupling is responsible for violation of the anticorrelation of
transmission and reflection fluctuations, $\delta T^2\neq - \,\delta R^2$, in
standard semiclassical approximation (SCA). Likewise, the standard SC
approximation is expected to fail for quantum shot noise \cite{ober} that is a
signature of this quantum indeterminism. For the special case of the
rectangular shaped dot with lead openings placed at the midpoints, summation
of exponentially proliferating pseudo-paths could be accomplished by tracing
trajectories in an extended zone scheme \cite{wir03,46}. For arbitrarily
shaped billiards and, in particular, chaotic billiards where already classical
trajectories proliferate exponentially, systematic inclusion  of paths and
pseudo-paths up to the same length is considerably more complicated. In the
following we address this problem within the framework of a semiclassical
Dyson equation. We present a diagrammatic expansion that allows a systematic
summation of classical path and pseudo-path contributions. Applications to the
circular billiard (as a prototypical regular system) and to the 
Bunimovich stadium
billiard (as the prototype system for chaotic scattering)
show good agreement with the
numerically calculated exact quantum path length spectrum. We furthermore
apply the PSCA to recent experimental studies of quantum scattering in
microwave billiards in which pseudo-path contributions could be experimentally
identified.

The plan of the paper is as follows: In section II we briefly review the
standard semiclassical approximation and previous attempts to include
diffraction effects. The pseudo-path semiclassical approximation for
arbitrarily convex shaped billiards will be presented in section III employing
a semiclassical version of the Dyson equation. We develop a diagrammatic
expansion of $G$ in terms of paths and pseudo-paths.  Its evaluation is
numerically facilitated by an algebraic matrix representation that reduces
path summations up to infinite order to a sequence of matrix multiplications
and
inversions, as discussed in section IV. Numerical results and comparison with
the full quantum results as well as microwave experiments are given in section
V,
followed by a short summary and outlook onto future applications in
sec. VI. Details on the Fraunhofer diffraction approximation are given in the
appendix.

\section{Standard semiclassical approximation}
 \begin{figure}[tb]
  \centering
  \includegraphics[draft=false,keepaspectratio=true,clip,%
                   width=0.95\linewidth]%
                   {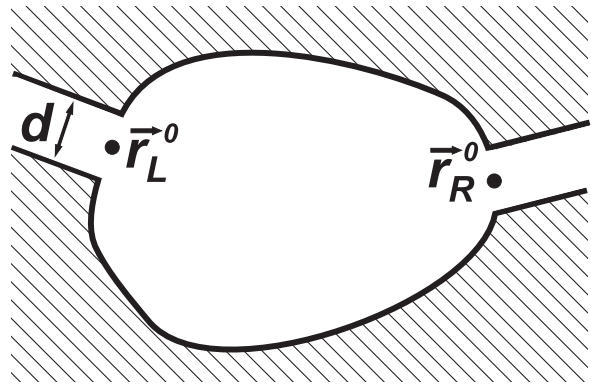}
 \caption[FIG1]{An open arbitrarily convex-shaped billiard with two narrow
   leads of equal width $d$: left $(L)$ and right $(R)$.}
 \label{opsquare}
 \end{figure}
The conductance of a ballistic two-terminal system (as depicted in Fig. 1) is determined by the scattering amplitudes $T_m$ through the Landauer formula \cite{land57}, 
\begin{equation}
 g (k) = \frac{2\, e^2}{h} \; \left( 
 \sum\limits_{m_L=1}^M\, \sum\limits_{m_R=1}^M\; 
 \vert T_{m_R,m_L} (k) \vert^2 \; \right) \; ,
 \label{ea:2-1}
\end{equation}
where $M$ is the number of open modes in the leads (quantum 
wires) 
and $T_{m_R, m_L}(k)$ are the transmission amplitudes from the $m^{th}$ mode 
in the entrance lead [referred to in the following as left (L) lead]  
to the $(m')^{th}$ mode in the exit lead 
[referred to as the right (R) lead]. In the following we choose local coordinate systems ($x_i,y_i$), 
$i=L,R$ for the leads, where $x_i$ denotes the longitudinal and $y_i$ the 
transverse direction of lead $i$. For simplicity we use the coding (L,R) for
the entrance and exit channels throughout this publication,
irrespective of the actual location at which the leads are attached. 
The projection of the retarded Green's propagator onto the transverse wave 
functions $\phi_{m_L} (y_L)$ and $\phi_{m_R} (y_R)$ of incoming and outgoing
modes
\cite{bar93} serves as starting point for most semiclassical theories which
approximate transmission amplitudes from mode $m_L$ to mode $m_R$:
 \begin{eqnarray}
 T_{m_L,m_R}\,(k)  & = & -i\sqrt{k_{x_R,m_R} k_{m_L,x_L}}\int dy_R \int dy_L\, 
 \phi_{m_R}^{*}(y_R) \nonumber \\
 && \times \,G\,(x_R,y_R,x_L,y_L,k)\,\phi_{m_L}\,(y_L)\; 
 \label{e:2-2}
 \end{eqnarray}
Here and in the following we use atomic units ($\hbar=|e|=m_{e}=1$).
The transverse wavefunctions $\phi_m(y)$ are given for zero magnetic field ($B=0$) by
 \begin{equation}
 \phi_m (y) = \sqrt{\frac{2}{d}} \; 
 \begin{cases}
 \cos \left( \frac{m\pi}{d} y \right) \qquad & \text{ $m$ odd}\\
 \sin \left( \frac{m \pi}{d} y \right) \qquad & \text{$m$ 
 even,}
 \end{cases}
 \label{e:2-3}
 \end{equation}
where $d$ denotes the width of the leads. We assume, for simplicity, that the two leads have identical width.
The scattering state in the lead is the product of the transverse lead 
wave function $\phi_m (y)$ and a plane wave in the longitudinal direction with 
wave vector $k_{x,m} = \sqrt{k^2 - \vert m\pi / d \vert^2}$. Both leads have a
total of $(m=1,\ldots,M)$ open (i.e.~transmitting) modes. Analogously, the
amplitude for reflection from the incoming mode $m_L$ into a different mode
$m_L'$ in the same quantum wire is given by 

\begin{eqnarray}
R_{m_L,m'_L} & = & \delta_{m_L,m'_L}- i\sqrt{k_{x_L, m_L'} k_{m_L, x_L}} \int{{dy'}_L}
\int{dy_L}\,
\nonumber \\
&& \phi^*_{m'_L}(y'_L) G (x'_L, y'_L, x_L, y_L, k) \phi_{m_L} (y_L) \, .
\label{equation:2-4}
\end{eqnarray}
Equations (\ref{e:2-2}) and (\ref{equation:2-4}) 
can be written in terms of flux 
normalized projectors onto the left (right) lead or point contact 
($x_{L,R}^0,y_{L,R}^0$) with matrix elements 
\begin{eqnarray}
\langle\vec{r}\phantom{,}|P_L|\vec{r}\phantom{,}'\rangle & = & \sqrt{|k_{x_L}| \; |k'_{x_L}|} \;\delta
(\vec{r}-\vec{r}\phantom{,}')\delta(x-x^0_L)  \\ 
&& \left[ \Theta(y-y^0_L+d/2) - \Theta 
  (y-y^0_L-d/2)\right] \nonumber
\label{e:2-5}
\end{eqnarray}

\begin{eqnarray}
\langle\vec{r}\phantom{,}|P_R|\vec{r}\phantom{,}'\rangle & = & \sqrt{|k_{x_R}| \; |k'_{x_R}|} \;  \delta(\vec{r}-
\vec{r}\phantom{,}') \delta(x-x^0_R) \\ 
&& \left[ \Theta(y-y^0_R+d/2) - \Theta
  (y-y^0_R-d/2)\right] \nonumber
\label{e:2-5b}
\end{eqnarray}
as 
\begin{subequations}
\label{e:2-6}
\begin{align}
\label{e:2-6a}
T_{m_R, m_L} (k) = - i\langle m_R|P_R \,G (k)\,
P_L|m_L\rangle
\end{align}
\begin{align}
\label{e:2-6b}
R_{m_L', m_L} (k)=\delta_{m_L,m'_L}- i\langle
m_L'|P_L\,G (k) \,P_L |
m_L\rangle \, .
\end{align}
\end{subequations}
The term $|k_{x_L}|$ ($|k'_{x_L}|$) denotes the longitudinal component
of the momentum of the incoming (outgoing) wave function at the left lead.

The semiclassical approximation to the scattering amplitudes is 
obtained by approximating the Green's function $G\, (\vec{r}_2,\vec{r}_1, k)$
in Eq.\ \eqref{e:2-2} by the semiclassical Green's propagator 
$G^{SC}\, (\vec{r}_2,\vec{r}_1, k)$.
The standard semiclassical Green's propagator $G^{SC} (\vec{r}_2, \vec{r}_1, k)$, the Fourier-Laplace transform of the van Vleck propagator evaluated in stationary phase approximation (SPA), 
describes the probability amplitude for 
propagation from $\vec{r}_1$ to $\vec{r}_2$ at a fixed energy, $E=k^2 /2$. 
It can be expressed in terms of a sum over all classical paths of 
energy $E$ (or, equivalently, wavevector $k$) connecting these two points 
\cite{gutz91a},
\begin{eqnarray}
G^{SC}\, (\vec{r}_2,\vec{r}_1, k)  &=&  
 \sum\limits_{\alpha :\,\vec{r}_1\to \vec{r}_2} G^{SC}_{\alpha} \; \nonumber \\ &=&
 \sum\limits_{\alpha :\,\vec{r}_1\to \vec{r}_2} 
 \frac{\vert D_{\alpha} (\vec{r}_2, \vec{r}_1, k) \vert^{1/2}}{(2\pi i)^{1/2}} \;\label{e:2.7} \\ && \times 
 \exp \left[\, i \left(S_{\alpha} ( \vec{r}_2, \vec{r}_1, k) - \frac{\pi}{2} \mu_{\alpha}\,\right) \right]. \nonumber
\end{eqnarray}
Here, $S_{\alpha} (\vec{r}_2, \vec{r}_1, k) = k L_{\alpha}$ is the action 
of the path $\alpha$ of length $L_\alpha$.
$D_{\alpha} (\vec{r}_2, \vec{r}_1, k)$
is the classical deflection factor \cite{gutz91a} which describes
the stability of the paths
and $\mu_{\alpha}$ denotes the Maslov index of the path $\alpha$. In line with the semiclassical approximation, 
the double integrals 
in Eqs.~\eqref{e:2-2} and (2.4) are frequently evaluated in stationary
phase approximation. 
Physically this means that the paths are entering and
exiting the cavity only with the discrete angles $\theta_m =
\mbox{arcsin}[m\pi/(dk)]$
due to the quantization of the transverse momentum in the leads.
For completeness we mention at this point that the path-sum in
Eq.~(\ref{e:2.7}) contains also those orbit pairs described in 
Ref.~\cite{ric94} which yield a weak-localization correction beyond the 
diagonal approximation.

Several strategies have been proposed to introduce diffractive effects in order
to quantitatively improve the semiclassical theory for transport through
open quantum billiards.
A straightforward way is to eliminate the SPA
for the double integral in Eq.~(2.4).
Expanding the action in the semiclassical Green's function Eq.~(\ref{e:2.7}) 
to first order in the transverse coordinate, the integral takes
the form of a Fraunhofer diffraction
integral and can be evaluated analytically \cite{wir97}.
On this level of approximation, diffraction effects are thereby automatically
included, however only upon entering and exiting the cavity, not during
propagation inside the cavity.
Schwieters et al. \cite{schw96} employed Kirchhoff diffraction theory
to calculate the diffractional weight of paths entering and exiting
the billiard. In addition, they introduced the concept of ``ghost-paths'':
paths that are specularly reflected at the lead opening due to 
diffractive effects.
The proper use of a diffractions weight for classical paths allowed
the quantitative determination of the peak-heights in the power spectra
of the transmission and reflection amplitudes \cite{schw96,wir97} - at
least for short path lengths. Ghost orbits could account for
some of the peaks that were missing in the semiclassical spectra
\cite{schw96}. Several deficiencies remained, however, unresolved: unitarity
of the semiclassical $S$ matrix is, typically, violated; the weak localization
peak is significantly underestimated, and the semiclassical pathlength
($\ell$) spectrum \cite{ish95},  

\begin{equation}
 P_{{m',m}}^{SC} (\ell) = \left\vert \int dk \; e^{ik\ell} \; T_{m',m}^{SC}
 (k) \; 
 \right\vert^2,
 \label{e:2.8}
\end{equation}

\noindent fails to account for all the peaks and overestimates the
 corresponding quantum pathlength spectrum for large $\ell$. For the special
 case of a rectangular billiard we could recently demonstrate \cite{wir03}
 that a systematic inclusion of pseudo-paths within the pseudo-path
 semiclassical approximation (PSCA) has the potential to overcome these
 deficiencies. In the following we derive the PSCA for an arbitrary convex
 billiard from a semiclassical Dyson equation and investigate its properties
 numerically. 

\section{Dyson equation for the PSCA}
Starting point is a semiclassical version of the Dyson equation for the
Green's function where the standard semiclassical Green's function $G^{SC}$
plays the role of the unperturbed Green's function and the diffractive
scatterings at the lead openings (or point contacts) are the perturbation. 
Accordingly, we have 
\begin{equation}
\label{e:3.1}
G^{PSC} = G^{SC} + G^{SC} V G^{PSC},
\end{equation}
where the perturbation ``potential'' $V$ is given in terms of the projectors
Eqs.~(2.5, 2.6) as
\begin{equation}
\label{e:3.2}
V=P_L+P_R \, .
\end{equation}
Iterative solution by summation 
\begin{equation}
\label{e:3.3}
G^{PSC}=G^{SC}\sum_{i=0}^{\infty}(VG^{SC})^i=G^{SC} \sum^{\infty}_{i=0} \left( (P_L+P_R) G^{SC} \right)^i
\end{equation}
includes diffractive scatterings into the $G^{PSC}$ to all orders.

The key to the multiple diffractive scattering expansion is that within the
semiclassical expansion each projection operator onto the $L$ and $R$ point
contacts selects classical trajectories emanating from or ending up at the
leads and at the same time spawns new generations of classical trajectories. 

Noting that $G^{PSC}$ will only be evaluated in the domain of $P_L$ or $P_R$,
Eq.\ (\ref{e:3.3}) can be reorganized in terms of a 2 x 2 matrix Dyson
equation. We decompose $G^{SC} $ as follows
\begin{subequations}
\label{e:3.4}
\begin{align}
\label{e:3.4a}
P_L \, G^{SC} P_L \Rightarrow G_{LL}^{SC} = \sum_{\alpha_{LL}} G_{\alpha_{LL}}^{SC}
\end{align}
\begin{align}
\label{e:3.4b}
P_L\, G^{SC} P_R \Rightarrow G_{LR}^{SC} = \sum_{\alpha_{LR}} G_{\alpha_{LR}}^{SC}
\end{align}
\begin{align}
\label{e:3.4c}
P_R \, G^{SC} P_L \Rightarrow G_{RL}^{SC}= \sum_{\alpha_{RL}} G_{\alpha_{RL}}^{SC}
\end{align}
\begin{align}
\label{e:3.4d}
P_R \, G^{SC} P_R \Rightarrow G_{RR}^{SC} = \sum_{\alpha_{RR}} G_{\alpha_{RR}}^{SC} 
\end{align}
\end{subequations}
 \begin{figure}[tb]
  \centering
  \includegraphics[draft=false,keepaspectratio=true,clip,%
                   width=0.95\linewidth]%
                   {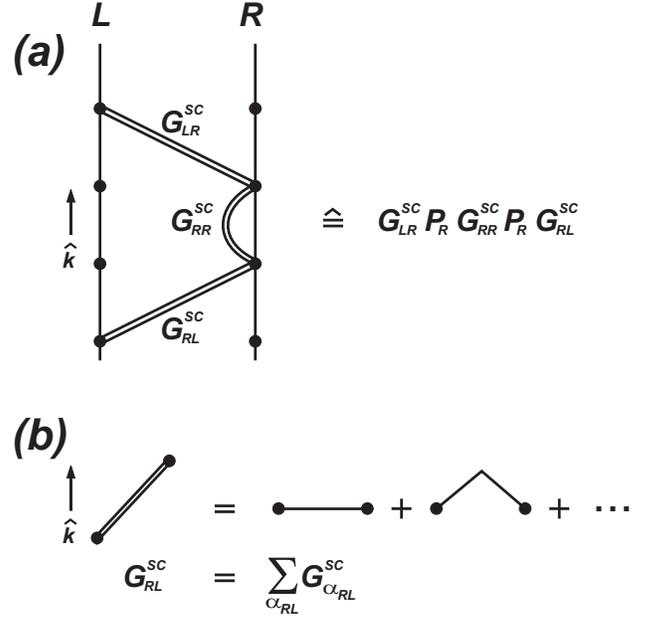}
 \caption[FIG2]{Diagramatic expansion of paths. Double line: $G^{SC}_{ij}$,
   single line: contribution of individual classical trajectories
   $G^{SC}_{\alpha_{ij}}$, dots: replica of point contacts along the propagation
   direction $\hat{k}$ (``time''). (a) Typical term appearing in the expansion
   of $G^{PSC}$. (b) Expansion of $G^{SC}_{ij}$ in terms of classical path
   contributions.}
  
 \label{pathexamples}
 \end{figure}
In Eq.\ (\ref{e:3.4}) the index $\alpha_{ij}$ refers to paths that emanate
from the point contact  $j \, (j=L,R)$ and end up at point contact $i \, (i =
L,R)$. Each of the four disjunct subsets of classical paths is, in general,
infinite. In the following we denote the truncated number of trajectories of
the corresponding class by $N_{ij}$. Each $G_{ij}^{SC}$ containing a large
finite number $N_{ij}$ or an infinite number of trajectories is
diagramatically represented by a double line, each contribution of an
individual trajectory by a single line (Fig.\ 2). The trajectories emanate or
end on vertices representing the $L$ or $R$ point contacts. 

In this formalism Eq.~(\ref{e:3.1}) now becomes
\begin{eqnarray}
\label{e:3.5}
{G_{LL}^{PSC} \,\, G_{LR}^{PSC} \choose G_{RL}^{PSC}
 \,\, G_{RR}^{PSC}} & = & {G_{LL}^{SC} \,\, G_{LR}^{SC} \choose G_{RL}^{SC} \,\, G_{RR}^{SC}} + {G_{LL}^{SC}  \,\, G_{LR}^{SC} \choose G_{RL}^{SC} \,\, G _{RR}^{SC}} \nonumber \\ & \times &
{P_L \,\, 0 \choose 0 \,\, P_R}
{G_{LL}^{PSC} \,\, G_{LR}^{PSC} \choose G_{RL}^{PSC} \,\, G_{RR}^{PSC}}
\end{eqnarray}
Its solution $G^{PSC}$ denoted by solid lines can now be diagramatically
represented (Fig.\ 3) as a sum over all pseudo-paths that result from the
couplings of classical-path Green's functions by successive diffractive
scatterings at point contacts. Only pseudopath combinations (or diagrams)
contribute that are connected at vertices $L$ or $R$.

Equation (\ref{e:3.5}) can be formally solved by matrix inversion,
\begin{eqnarray}
\label{e:3.6}
{G_{LL}^{PSC} \, \, G_{LR}^{PSC} \choose
G_{RL}^{PSC} \, \, G_{RR}^{PSC}} &=&
{1 - G_{LL}^{SC} P_L \, \, \quad- G_{LR}^{SC} P_R\,  \choose
\,- G_{RL}^{SC} P_L  \, \,\quad 1 - G_{RR}^{SC} P_R}^{-1} \nonumber \\ & \times & 
{G_{LL}^{SC} \, \, G_{LR}^{SC}  \choose
G_{RL}^{SC} \, \, G_{RR}^{SC}}\,,
\end{eqnarray}
resulting in a sum over (up to) infinitely long pseudo
paths with (up to) an infinite number of diffractive scatterings. Note that
the number of classical paths between two diffractive ``kinks''  and their
lengths may reach infinity as well. The handling of this double limit plays an
important role in the numerical implementation as destructive interference of pseudo-paths and classical paths of
comparable length must be properly taken into account. 
The evaluation of Eq.~(\ref{e:3.5}) and (\ref{e:3.6}) is not straight-forward
as each operator product in Eq.~(\ref{e:3.3}) contains a multi-dimensional
integral over  $I\!\!R^2$. For illustrative purposes we explicitly give the
first-correction term ($ G^{SC} P_L G^{SC} + G^{SC} P_R G^{SC}$) of
Eq.~(\ref{e:3.5}). We have to calculate the double integral e.g., 
\begin{eqnarray}
\label{e:3.7}
G^{SC}_{LR} P_R G^{SC}_{RL} &=& \int \limits_{I\!\!R^2} d^2r' \int \limits_{I\!\!R^2} d^2r \;  G^{SC}_{LR}(\vec{r}_2,\vec{r}\phantom{.}') \;
 P_R(\vec{r}\phantom{.}',\vec{r}\phantom{.}) \; \nonumber \\ & \times & G^{SC}_{RL}(\vec{r},\vec{r}_1)\,,
\end{eqnarray}
where $\vec{r}_2\,(\vec{r}_1)$ are both located on the left point contact.
The double integral reduces to a one-dimensional integral along the lead openings due to the $\delta$-functions in the projector Eq.~(\ref{e:2-5}),
\begin{eqnarray}
\label{e:3.8}
G^{SC}_{LR} P_R G^{SC}_{RL} && =  \int_{-d/2}^{d/2} dy \;\sqrt{|k_{x_R}| \; |k'_{x_R}|} \;
 \\  \times &&
G^{SC}_{LR} (\vec{r}_2,x^0_R,y^0_R+y) \;
G^{SC}_{RL} (x^0_R,y^0_R+y,\vec{r}_1). \nonumber
\end{eqnarray}
Inserting the expression for the semiclassical Green's function
Eq.~(\ref{e:2.7}), and using the abbreviation $\vec{r}_R=(x^0_{R},y^0_R+y)$ we
obtain
\begin{widetext}
\begin{eqnarray}
G^{SC}_{LR} P_R G^{SC}_{RL} & = & 
- \sum_{\alpha_{RL}} 
  \sum_{\alpha_{LR}} \;
  \int\limits_{-d/2}^{d/2} 
  dy \;  \sqrt{|k_{x_R}| \; |k'_{x_R}|} \; G^{SC}_{\alpha_{LR}}(\vec{r}_2,\vec{r}_R) \; G^{SC}_{\alpha_{RL}} (\vec{r}_R,\vec{r}_1)
 =  - \frac{1}{2\pi i}  
 \sum_{\alpha_{RL}} 
 \sum_{\alpha_{LR}} 
 \left|\bar{D}_{\alpha_{RL}}| \right.
 \left|\bar{D}_{\alpha_{LR}}| \right. 
  \nonumber \\
& & \times \int\limits_{-d/2}^{d/2} dy \;  
\sqrt{|k_{x_R}| \; |k'_{x_R}|}
 \exp\left[i k \left(L_{\alpha_{LR}}(\vec{r}_1,\vec{r}_R)+
                L_{\alpha_{RL}}(\vec{r}_R,\vec{r}_2)\right)
          -i\frac{\pi}{2}\left(\mu_{\alpha_{RL}}+\mu_{\alpha_{LR}}\right) \right] \, , \phantom{xx} 
\label{hallo1}          
\end{eqnarray}
\end{widetext}
where we have assumed that the classical deflection factors 
$D_\alpha$ are smooth functions over the range of the lead mouth and
can be approximated by their value $\bar{D}_\alpha$ at the center. In Eq.~(\ref{hallo1}) $\, \alpha_{RL}$ denotes paths connecting the left with the right lead ($L \rightarrow R$) while $\alpha_{LR}$ represents paths $(R \rightarrow L)$.
 \begin{figure}[b]
  \centering
  \includegraphics[draft=false,keepaspectratio=true,clip,%
                   width=1.0\linewidth]%
                  {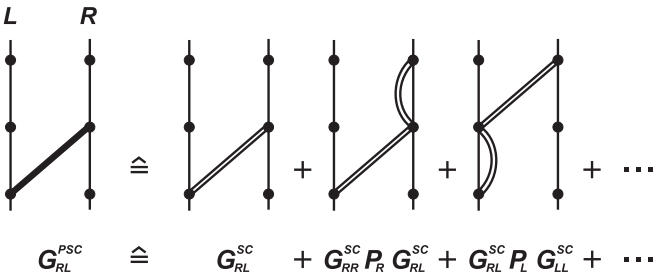}

 \caption[FIG3]{Thick solid line: $G^{PSC}$; first terms of the expansion of the Dyson equation for $G^{PSC}_{RL}$.}
 \end{figure}
In order to solve the integral in Eq.~(\ref{hallo1}) analytically,
we expand the path length $L_{\alpha}$  of the classical paths 
$\alpha$ leading from point $\vec{r}_i$ to the lead mouth $\vec{r}_R$
(here we set $y_R=0$ for simplicity and only consider regular billiards):
\begin{equation}
L_{\alpha_{RL}}(\vec{r}_R,\vec{r}_i) = L_{\alpha_{RL}} + \sin(\theta_{\alpha_{RL}}) y
 + \frac{\cos(\theta_{\alpha_{RL}})}{L_{\alpha_{RL}}} y^2 \hdots,
\label{lexpand}
\end{equation}
where $L_{\alpha_{RL}} = L_{\alpha_{RL}}(\vec{r}_R, \vec{r}_i)$ is the length
of the path that reaches the center of the lead mouth. 

Taking into account only the first-order correction in the lateral displacement, the integral in Eq.~(\ref{hallo1})
takes the form of a Fraunhofer diffraction integral and can be solved
analytically \cite{wir97,wir03} (see also appendix),
\begin{eqnarray}
I &=& \int\limits_{-d/2}^{d/2} dy  
\sqrt{|k_{x_R}(y)| \; |k'_{x_R}(y)|}  \, \nonumber \\ && \times 
\exp \left\{i k \left[L_{\alpha_{LR}}(y)+L_{\alpha_{RL}}(y)\right]\right\}\\
& \approx & k \sqrt{ \cos( \theta_{\alpha_{LR}}) \cos( \theta'_{\alpha_{RL}}) }
 \, \int\limits_{-d/2}^{d/2} dy 
\exp\left\{i k \left[L_{\alpha_{LR}}+ L_{\alpha_{RL}}\right] \right. \nonumber \\
&& + 
 \left.
ik \left[\sin(\theta_{\alpha_{LR}}) + \sin(\theta '_{\alpha_{RL}})\right]y
\right\} \nonumber \\
& = & \exp \left[i k  \left(L_{\alpha_{LR}}+ L_{\alpha_{RL}} \right) \right] 
r(\theta_{\alpha_{LR}},\theta'_{\alpha_{RL}},k) ,\nonumber
\label{e:3.11}
\end{eqnarray}
where $\theta'_{\alpha_{RL}}$ is the ending angle of the incoming path and $\theta_{\alpha_{LR}}$ 
is the starting angle of the exiting path. 
Furthermore, $r(\theta_{\alpha_{LR}},\theta'_{\alpha_{RL}},k)$ corresponds 
to the Fraunhofer reflection coefficient at an open lead \cite{wir03}:
 
\begin{eqnarray}
\label{e:3.12}
r(\theta_{\alpha_{LR}},\theta'_{\alpha_{RL}},k)   & = & 
2\sqrt{ \cos( \theta_{\alpha_{LR}}) \cos( \theta'_{\alpha_{RL}}) }\, \\ & \times &
\left\{\frac{ \sin\left[ \frac{kd}{2}
(\sin  \theta_{\alpha_{LR}} + \sin  \theta'_{\alpha_{RL}}) \right]}
{\sin  \theta_{\alpha_{LR}} + \sin  \theta'_{\alpha_{RL}} }\right\} \nonumber
\end{eqnarray}
\vspace{2ex}

\noindent Finally, this first order correction term in Eq.~(\ref{e:3.7})
can be written explicitly as
\begin{equation}
G_{LR}^{SC} \, P_R \, G_{RL}^{SC} =
\sum_{\alpha_{LR}}  
\sum_{\alpha_{RL}} 
G_{\alpha_{LR}}^{SC} r(\theta_{\alpha_{LR}},\theta'_{\alpha_{RL}},k)
G_{\alpha_{RL}}^{SC} \, .
\label{firstcorr}
\end{equation}

 \noindent Note that classical paths contributing
to $G^{SC}$ do not incorporate the finite width of the point contact since we take the limit $ d \rightarrow 0$ in the simulation of classical paths. Instead they are
specularly reflected at the open lead mouth $R$ as if there
were a hard wall.
Such a path will interfere with the pseudo-path  that has almost the
same length and topology  but experiences a non-specular, i.e.~diffractive
reflection at the same point contact.

All higher-order corrections are evaluated analogously. Each additional vertex
connecting two $G_{SC}$ gives rise to an additional Fraunhofer integral with
an interior reflection amplitude $r(\theta', \theta)$.

\section{Matrix representation of pseudo-path sum}
A numerical representation of the Dyson equation
[Eqs.~(\ref{e:3.5},\ref{e:3.6})] requires the truncation of the number of
contributing paths in each of the four basic Green's functions
[Eq.~(\ref{e:3.4})] to large but finite numbers. Let $N_{LL}$ denote
the number of paths leading from $\vec{r}_L^0$ back to $\vec{r}_L^0$.
Accordingly, $N_{RR}$ is the number of paths of the class $(R \rightarrow R)$,
$N_{RL}$ the number for $(L \rightarrow R)$ and $N_{LR}$ the number
for $(R \rightarrow L)$. Note that $N_{RL} = N_{LR}$ for systems with
time reversal symmetry.
The total number of contributing paths is 
$N = N_{LL} + N_{RR} + N_{LR} + N_{LR}$.
In order to numerically solve the semiclassical Dyson Equation (\ref{e:3.5}),
we write the semiclassical Green's function $G^{SC}$ as a diagonal 
$N \times N$ matrix where each diagonal matrix element represents the
contribution of one particular classical path.
Distinguishing the four different subclasses of paths,
the matrix can be written in the following form:
\begin{equation}
\label{e:4.3}
\underline{\underline{G}}^{SC} (k) =
\left( 
\begin{array}{cccc}
\underline{\underline{G}}_{LL}^{SC}(k) & & & 0\\
& \underline{\underline{G}}_{RL}^{SC}(k) & &  \\
  & & \underline{\underline{G}}_{RR}^{SC}(k)& \\
0& & & \underline{\underline{G}}_{LR}^{SC}(k) 
\end{array}\right) \, ,
\end{equation}
where the matrix elements of the sub-matrices are given by, e.g., 
\begin{equation}
 \underline{\underline{G}}_{\alpha_{LL}}^{SC}(k) =
 \frac{\vert D_{\alpha_{LL}} \vert^{1/2}}{(2\pi i)^{1/2}} 
 \exp \left[\, i \left(kL_{\alpha_{LL}} - \frac{\pi}{2} \mu_{\alpha_{LL}}\,\right) \right],
\label{e:bla}
\end{equation}
and $\alpha_{LL} = 1,\ldots,N_{LL}$.

Due to diffractive coupling at the lead mouths, the matrix 
$\underline{\underline{G}}^{PSC}$ that represents the 
Green's function in the pseudo-path semiclassical approximation
contains also non-diagonal matrix elements.
A specific matrix element $(\alpha',\alpha)$ consists 
of a sum of all path-combinations that contain the classical
path $\alpha$ as first segment and the classical paths $\alpha'$ as
last segment.

The coupling between different classical paths is represented by 
the non-diagonal vertex matrix $\underline{\underline{r}} (k)$.
The matrix elements are given - within the Fraunhofer 
diffraction approximation - by the
reflection amplitudes [Eq.~(\ref{e:3.12})]. Since
$r(\theta, \theta',k)$ depends on both the angle of emission, $\theta$, of
the new path and the angle of incidence, $\theta'$, of the previous path,
16 block matrices would result. However due to the restriction
imposed by the projections $P_L $ and $P_R$ that the endpoint ($L$ or $R$) of
the incoming trajectory must agree with the starting point of the outgoing
trajectory, effectively only 8 block matrices can contribute. These can be
characterized by the starting points and endpoints of classical trajectories
at the vertex. The block matrices resulting from the projection $P_L$ are
$\underline{\underline{r}}_{LL,LL}, \,\,
\underline{\underline{r}}_{LL,LR},\,\,  \underline{\underline{r}}_{RL, LL}$
and $\underline{\underline{r}}_{RL,LR}$. Analogously, from the vertex $P_R,\;$
we get $ \underline{\underline{r}}_{RR,RR},\;
\underline{\underline{r}}_{RR,RL}, \; \underline{\underline{r}}_{LR,RR}, $ and
$\underline{\underline{r}}_{LR,RL}$. 
E.g., $\underline{\underline{r}}_{LL,LR}$ is a $N_{LL} \times N_{LR}$ matrix
with the elements
\begin{equation}
\label{e:4.1}
\underline{\underline{r}}_{\alpha_{LL},\alpha'_{LR}}(k) = 
\left\{ r \left( \theta_{\alpha_{LL}}, \theta'_{\alpha_{LR}}, k \right) \right\}.
\end{equation}

The full matrix is
\begin{equation}
\label{e:4.2}
\underline{\underline{r}} (k) = \left(
\begin{array}{cccc}
\underline{\underline{r}}_{LL,LL} (k) & 0 & 0 & \underline{\underline{r}}_{LL,LR}(k) \\
\underline{\underline{r}}_{RL,LL} (k) & 0 & 0 & \underline{\underline{r}}_{RL,LR} (k) \\
 0 & \underline{\underline{r}}_{RR,RL} (k) &  \underline{\underline{r}}_{RR,RR} (k)  & 0\\
 0 & \underline{\underline{r}}_{LR,RL} (k) & \underline{\underline{r}}_{LR,RR}(k) & 0 \\
\end{array}\right)\,.
\end{equation}
For billiards of arbitrary shape and positions of the leads, no
further reductions are possible. Only for structures with discrete geometric
symmetries $(L \leftrightarrow R)$ or time-reversal symmetry, the number of
non-equivalent trajectories and thus of independent amplitudes $r (\theta,
\theta',k)$ is reduced.

With Eqs.~(\ref{e:4.1}) and (\ref{e:4.2}), the Dyson equation
Eq.~(\ref{e:3.5}) can now be written as an algebraic matrix equation
\begin{subequations}
\label{e:4.4}
\begin{align}
\label{e:4.4a}
\underline{\underline{G}}^{PSC} (k) = \underline{\underline{G}}^{SC} (k) \sum^{\infty}_{i=0} \left[ \underline{\underline{r}} (k) \underline{\underline{G}}^{SC} (k) \right]^i 
\end{align}
\begin{align}
\label{e:4.4b}
\phantom{xxxxxxxx}= \underline{\underline{G}}^{SC}(k) \left[ 1 - \underline{\underline{r}} (k) \underline{\underline{G}}^{SC} (k)\right]^{-1}\,.
\end{align}
\end{subequations}

\noindent Equation (\ref{e:4.4b}) represents the ``exact'' summation over
pseudo-paths with up to an infinite number of diffractive scatterings. The 
accuracy of the result is, however, limited by the fact that we can only
take into account a finite number $N$ of classical paths. 

Finally, calculation of the $S$ matrix elements requires the projection of
$\underline{\underline{G}}^{PSC}$ onto the asymptotic scattering states in the
left and right quantum wire (Eqs.~(2.7,2.8). Following the same line of
reasoning as in Eqs.~(\ref{hallo1},\ref{lexpand},3.11,\ref{e:3.12}),
the projections $P_L$ and $P_R$ give rise to a transmission amplitude in
Fraunhofer diffraction approximation. The transmission amplitude from incoming
mode $m$ to a classical path $\alpha$ inside the cavity with launching angle
$\theta_\alpha$ is given by \cite{wir03}.

\begin{eqnarray}
t_m(\theta_\alpha,k) & = & \sqrt{\frac{2\cos\theta_\alpha}{kd}} \left[
\frac{\sin\left[\left(
k\sin\theta_\alpha + \frac{m\pi}{d} 
\right)\frac{d}{2}\right]}
{\sin\theta_\alpha + \frac{m\pi}{kd}} \right.  \nonumber
\\ +&& \left.
\frac{\sin\left[\left(
k\sin\theta_\alpha - \frac{m\pi}{d} 
\right)\frac{d}{2}\right]}
{\sin\theta_\alpha - \frac{m\pi}{kd}} 
\right].
\label{e:4.5}
\end{eqnarray}

Likewise, the transmission amplitude for a trajectory $\alpha$ approaching the point contact with angle $\theta_\alpha$ to exit in mode $m$ is also given by Eq.~(\ref{e:4.5}).

With Eq.~(\ref{e:4.5}) we form now amplitude matrices to map the asymptotic
scattering states onto the $N \times N$ representation of $G^{PSC}$. The $N
\times 2M$ matrix for the incoming state, where $M$ is the number of open
transverse modes $(m_{L,R}= 1,\ldots,M)$, is given by

\begin{equation}
\label{e:4.6}
\underline{\underline{A}}^{in} (k) = 
\left(
\begin{array}{cc}
t_{m_L}(\theta_{\alpha_{LL}}, k)& 0\\
t_{m_L} (\theta_{\alpha_{RL}}, k) & 0\\
0 & t_{m_R}(\theta_{\alpha_{RR}}, k)\\
0 & t_{m_R}(\theta_{\alpha_{LR}}, k) \\
\end{array}
\right)
\end{equation}

\noindent The corresponding projection amplitude for the outgoing scattering state is a $2M \times 2N$ matrix and reads
\begin{widetext}
\begin{equation}
\label{e:4.7}
\underline{\underline{A}}^{out}(k)=
\left(
\begin{array}{cccc}
t_{m_{L}} (\theta_{\alpha_{LL}}, k) & 0 & 0 & t_{m_{L}} (\theta_{\alpha_{LR}}, k) \\
0 & t_{m_R} ( \theta_{\alpha_{RL}}, k) & t_{m_R} (\theta_{\alpha_{RR}}, k) & 0 \\
\end{array}
\right).
\end{equation}
\end{widetext}
The $2M \times 2M$ dimensional $S$ matrix follows now from

\begin{equation}
\label{e:4.8}
\underline{\underline{S}}^{PSC} (k)  = \underline{\underline{A}}^{out}(k) \, \underline{\underline{G}}^{PSC} (k) \, \underline{\underline{A}}^{in} (k)
\end{equation}
\begin{equation}
\nonumber
\phantom{xxxxxxx}= {R_{m'_{L}, m_L} \qquad T_{m'_{L}, m_R} \choose
T_{m'_{R}, m_L} \qquad  R_{m'_{R}, m_R }}\, .
\end{equation}

\noindent With Eq.~(\ref{e:4.8}) the semiclassical calculation of the $S$
matrix of a hard-walled quantum billiard with two point contacts is reduced to
a sequence of matrix multiplications. System-specific input are the data
(length, angle of incidence and emission from point contact and Maslov index)
of each of the four classes $(L \rightarrow L, L \rightarrow R, R \rightarrow L, R
\rightarrow L)$ of classical trajectories.

\section{Numerical results}
In the following we will apply the pseudo-path semiclassical approximation to
the calculation of transmission and reflection amplitudes in different regular
and chaotic structures (i.e.~circle, rectangle and stadium). We compare the
PSCA calculations with the results of the standard semiclassical approximation
\cite{wir97}, exact quantum calculations \cite{MRGM} and microwave
experiments \cite{blo02}.

In order to evaluate the scattering matrix elements $T_{m',m}$ and $R_{m',m}$
numerically we have used the truncated form of Eq.~(\ref{e:4.4a}) with $i \leq
K$, where $K$ is the maximum number of diffractive scatterings (or
``kinks''). We therefore relate  $K$ to the maximum path-length $\ell_{max}$
of classical paths included. In order to include all pseudo-paths  with 
at least the
same length, we require that the length of the shortest trajectory in the
system $\ell_{min}$ multiplied by $K$ + 1 has to be larger than $\ell_{max}$
to guarantee that all pseudo-paths  up to  $\ell_{max}$ are included. Hence we
require

\begin{equation}
K \geq \left[\; \frac{\ell_{max}}{\ell_{min}}\; \right]\; - 1,
\label{e8.2.2}
\end{equation}
where the bracket stands for the largest integer less than the argument.
This requirement only assure that each classical path is shadowed by a
pseudo-path of comparable length. The converse is evidently not the
case. Pseudo-paths are permitted with path-length up to $\ell_{max} (K + 1)$
for which no classical counterpart of comparable length is included. We
account for this deficiency by Fourier filtering the power spectra $P(\ell)$
of the $S$-matrix elements for $\ell > \ell_{max}$. The classical input data
are determined analytically for the square and circle billiards while for the
chaotic stadium billiards the data are generated by calculating the
classical trajectories numerically.\\

\noindent {\bf A \,\,  Regular structures}\\
For regular structures classical and pseudo-paths can be easily enumerated and
calculations up to very high path length $(\ell=40)$ have been performed as a
sensitive test of semiclassical approximations. 
We present in the following
results for the path-length power spectrum [Eq.~(\ref{e:2.8})]. Figs.~4 and 5
show the power spectra of the transmission and reflection amplitudes for the
second mode, $T_{22}$ and $R_{22}$, in the circular billiard with
perpendicular leads. Note that all the scattering
geometries which we
investigate in the following have the same cavity area $A=4+\pi$, or,
accordingly, a linear dimension $D=\sqrt{4+\pi}\approx 2.7$\,.

Here we mainly concentrate on the improvement of the PSCA in comparison to the
SCA. For $T_{22}$, the overestimation of long paths ($\ell > 18$) in the SCA
calculation is clearly visible: compare the peaks labeled by black arrows in Fig.~4a with those in Fig.~4b.
 \begin{figure}[tb]
  \centering
  \includegraphics[draft=false,keepaspectratio=true,clip,%
                   width=0.95\linewidth]%
                  {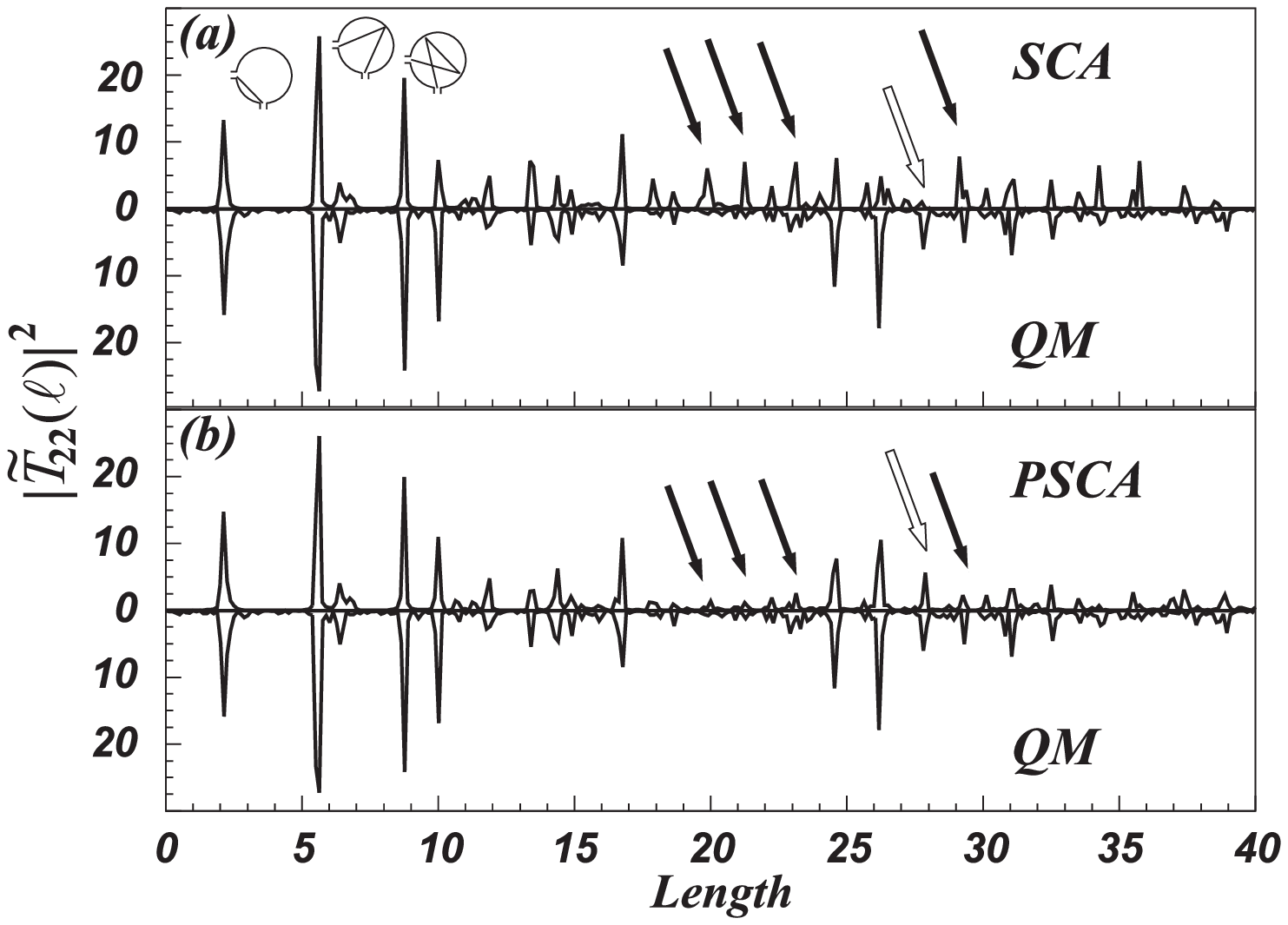}
 \caption[FIG4]{Power spectra $|\tilde{T}_{22}(\ell)|^2$ of the transmission
   amplitude in the circle billiard with perpendicular leads with
$R=\sqrt{1+4/\pi}$ and $d=0.25$ (see inset Fig.~5) for a finite window of $k$, $1 \le k \le 6$ in
units of $\pi/d$. } 
  \includegraphics[draft=false,keepaspectratio=true,clip,%
                  width=0.95\linewidth]%
                  {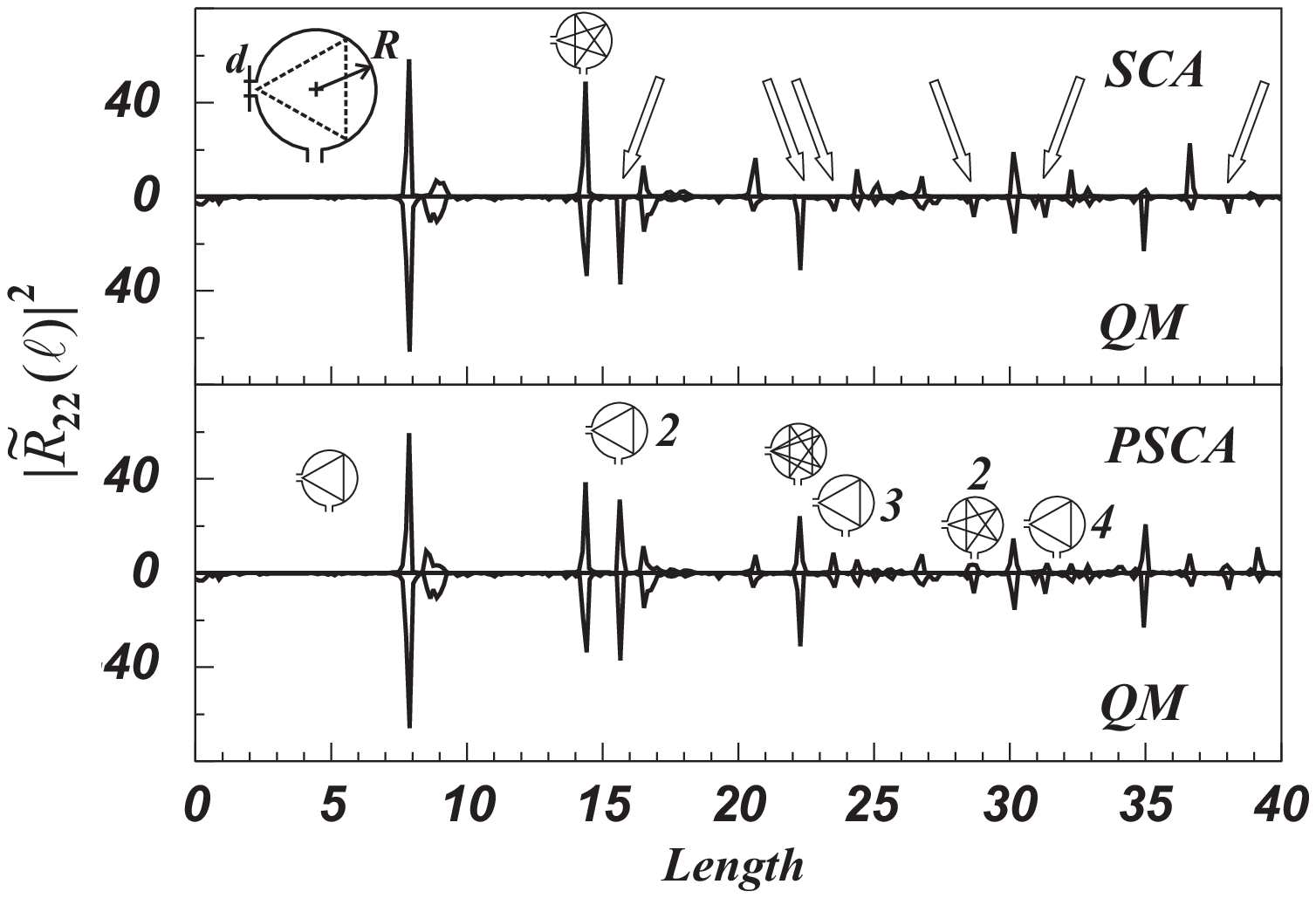}
 \caption[FIG5]{Power spectra $|\tilde{R}_{22}(\ell)|^2$ of the transmission
   amplitude in the circle billiard with perpendicular leads with
$R=\sqrt{1+4/\pi}$ and $d=0.25$ for a finite window of $k$, $1 \le k \le 6$ in
units of $\pi/d$. For
   more details see text.} 
 \end{figure}
The contributions of the additional
pseudo-paths included in the PSCA are responsible for the cancellation effects
which suppress the peak heights. The peak heights of the PSCA  agree well with
the quantum calculations for long paths. Also the lack of individual peaks
(i.e.~pseudo-paths) present in the quantum spectrum but missing in the SCA
spectra (e.g. at $\ell \approx 27.5$; see white arrows in Fig.~4) can be addressed in the transmission amplitude. 
 
This defect of the SCA can be seen more clearly in the plot for $R_{22}$ (see
arrows in the upper part of Fig.~5). The pseudo-path semiclassics provides for
the additional peaks in agreement with the quantum calculation. The insets in
the lower part of Fig.~5 show schematically the geometry of dominant
pseudo-paths, which experience at least one diffractive reflection at an open
lead. The number next to the insets gives the information on how often the
segment of a classical path has to be traversed to build up the corresponding
pseudo-path.\\

 \begin{figure}[tb]
  \centering
  \includegraphics[draft=false,keepaspectratio=true,clip,%
                   width=0.95\linewidth]%
                   {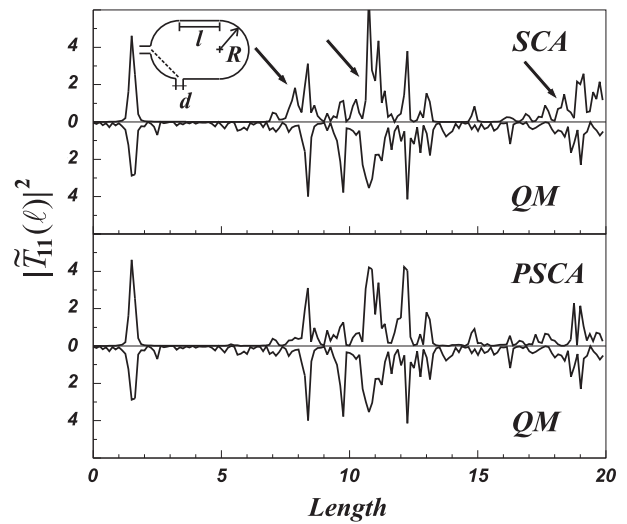}
 \caption[FIG6]{Power spectra of the transmission amplitude ($m=n=1$) in the
   stadium billiard with perpendicular leads
  with $R=1$, $l=2R$, and $d=0.25$ for a finite
 window of $k$, $0 \le k\le5$ in units of $\pi /d$.} 
 \end{figure}
\noindent {\bf B\, \, Chaotic structures (the Bunimovich stadium)}\\
Up to now we have compared the SCA and PSCA for open billiards with regular
classical dynamics. We turn now to chaotic (i.e.~non-regular) billiards. The
Bunimovich stadium \cite{bene78,bog84,rott00}
serves as prototype system for structures with chaotic classical
dynamics. Before we discuss our numerical results, we point out some of the
characteristic differences between the regular and chaotic structures. 
In the stadium billiard the number of path bundles \cite{wir97} up to a fixed length increases exponentially. The exponential proliferation represents a major  challenge for the summation of paths. This leads to technical problems for the calculation of
transport quantities in chaotic structures \cite{rott00} due to the limitation of the number of
paths that realistically can be taken into account. Furthermore, the path length distribution in chaotic structures differs qualitatively 
from that of regular structures. For the latter we find an algebraic
decay for the classical path-length distribution in contrast to the
exponential decay for classically chaotic
structures. This is to be distinguished from the exponential decay behavior
for pseudo-paths. Wirtz et al. \cite{wir03} have shown that for the
rectangular billiard 
the SCA leads to a linear (algebraic) decay of the path length power spectra,
while the PSCA
 gives rise to an exponential decay when pseudo-paths are included. However, in
chaotic systems, already classical paths proliferate exponentially 
as a function of the path length
and can account for exponential suppression of large 
path lengths. Therefore, the lack of pseudo-paths which also proliferate 
exponentially  is less dramatically felt than in regular 
systems where the exponential proliferation of pseudo-paths 
competes with only linear proliferation of classical paths.
This observation is key to the surprising findings in previous 
semiclassical calculations that the agreement between the 
semiclassical and the quantum pathlength spectrum is  better 
for chaotic rather than for regular systems 
\cite{rott00,wir97,bar93,ish95,naz01}. However, also in chaotic systems, the effects of diffractive pseudo-paths can be clearly seen. Fig.~6 shows the power spectra of transmission amplitudes for the stadium with perpendicular leads. For $T_{11}$, remarkably, the path-length spectra within PSCA displays {\it fewer} pronounced peaks than the SCA in agreement with the quantum calculation. The reason is that in a chaotic system the high density of pseudo-paths effectively causes path shadowing of true classical paths. As a result, some of the classical peaks are drastically reduced by destructive interference even for comparatively short path length.\\

\noindent {\bf C \, \, Comparison with microwave experiments}\\
As a third application we discuss the comparison between experimental studies of 
geometry-specific quantum scattering in microwave billiards \cite{blo02} and
semiclassical approximations (SCA and PSCA). The physics and modeling
of microwave cavities are conceptually similar to that of semiconductor quantum
dots due to the equivalence of the time-independent Schr\"odinger and
Helmholtz equations \cite{kim02}.
Moreover, for microwave frequencies $\nu < \nu_{max}=c/2h$ \cite{kim02}, where
$h$ is the height of the microwave billiard, only a single transverse mode is
supported by the cavity. This reduces the electromagnetic boundary
conditions to Dirichlet-boundary conditions
 allowing for an exact correspondence between electrodynamics and quantum
 mechanics. Accordingly, the component of the electrical field perpendicular
 to the plane of the microwave billiard corresponds to the quantum mechanical
 wave function.

A direct measurement of transmission and reflection amplitudes (i.e.~electrical field amplitudes) becomes possible. 
Moreover the semiclassical analysis presented above is directly applicable to describe
scattering in microwave billiards.
  \begin{figure}[tb]
  \centering
  \includegraphics[draft=false,keepaspectratio=true,clip,%
                   width=0.95\linewidth]%
                   {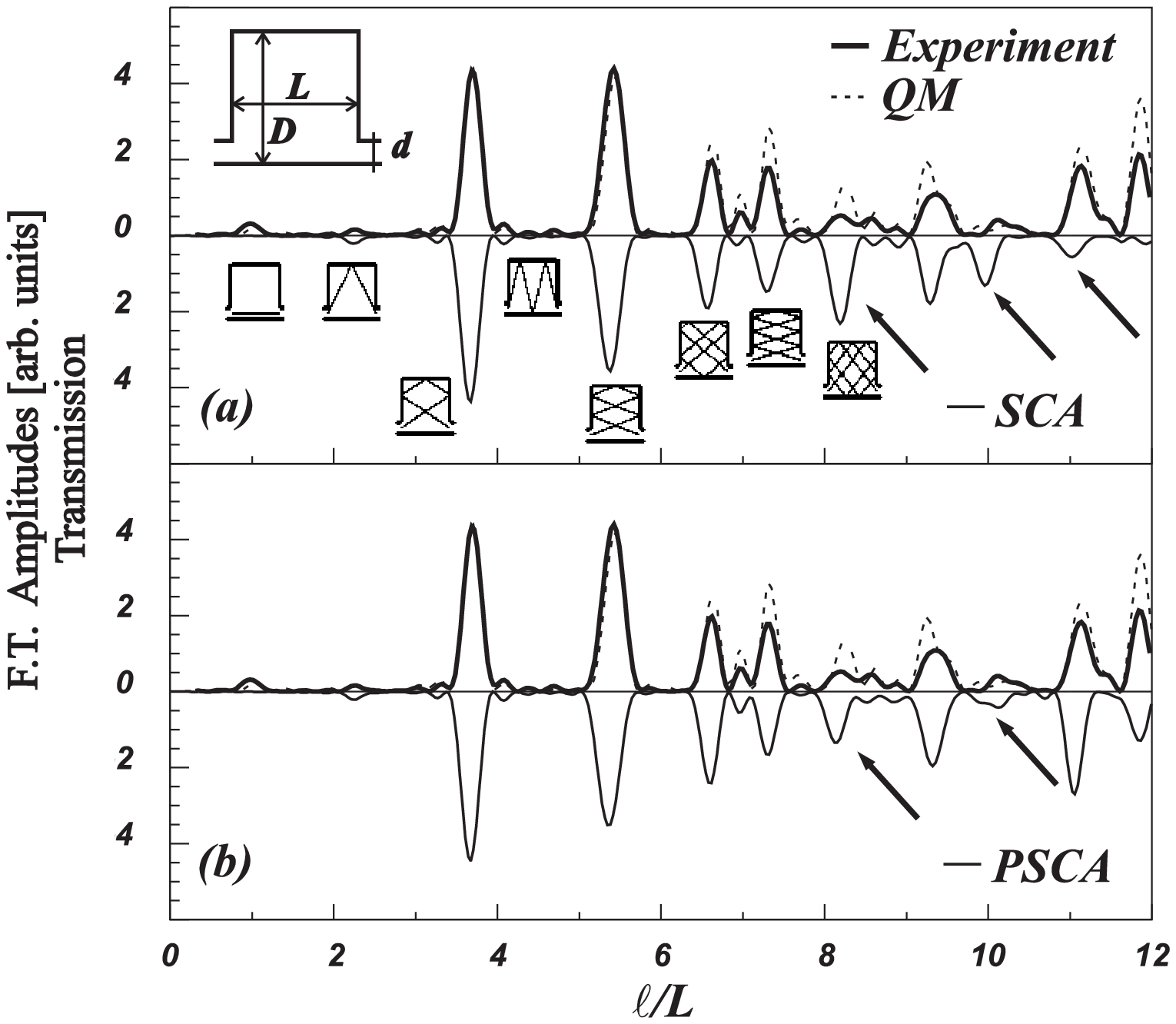}
 \caption[FIG7]{
 Power spectra $|\tilde{T}_{11}(\ell)|^2$ of the experimental and calculated
 (QM, SCA and PSCA) transmission  amplitude for the rectangular billiard
 ($L=225$mm, $D=237$mm and $d=15.8$mm)  with opposite leads (not centered)
for a finite window of $\nu$, $13\,$GHz $ \le \nu \le 18\,$GHz. The
 characteristic peaks are identified in terms of classical transmitted
 trajectories (insets). Experimental data by H.~Schanze \cite{blo02}.} 
 
  \centering
  \includegraphics[draft=false,keepaspectratio=true,clip,%
                   width=0.95\linewidth]%
                   {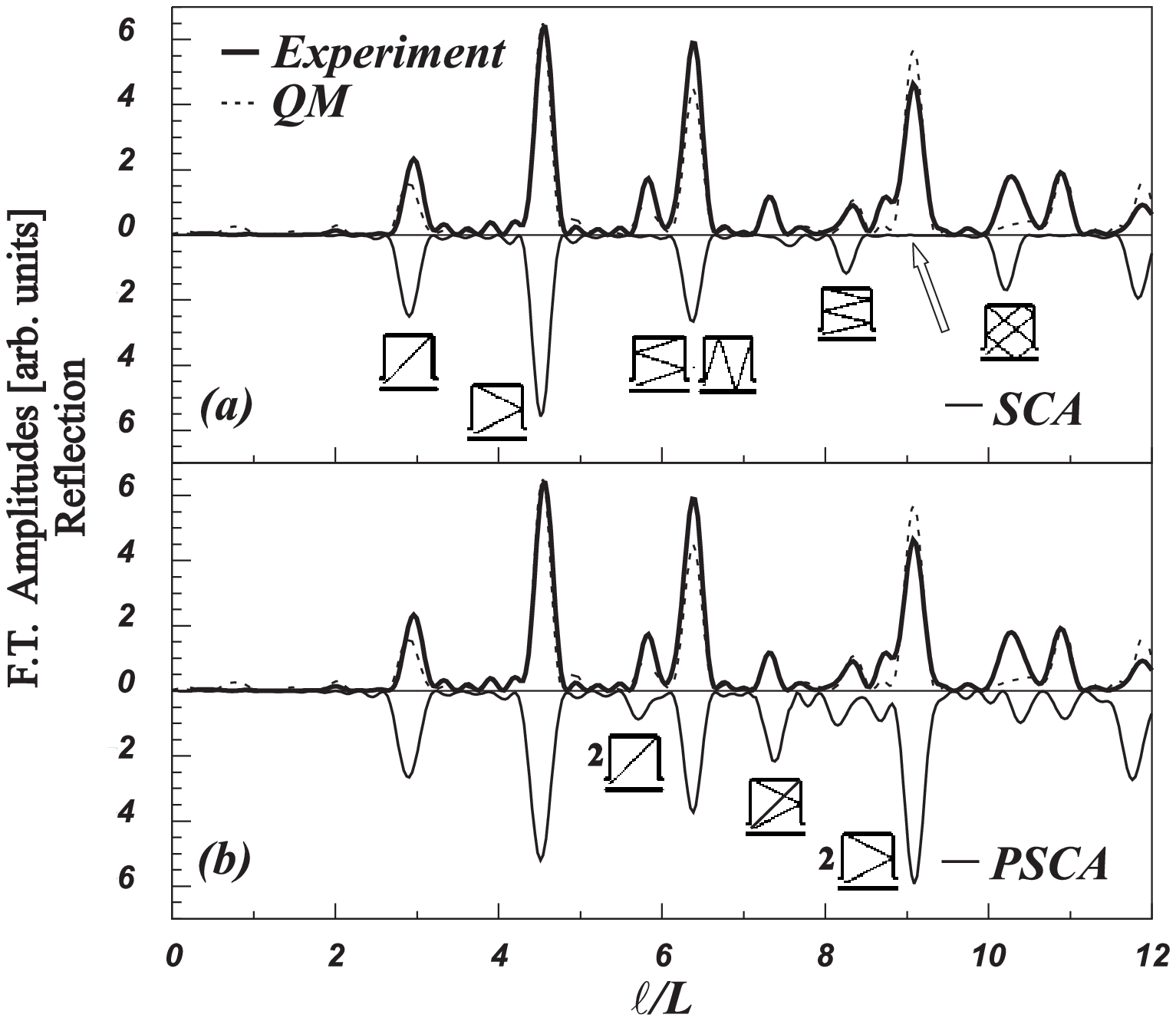}
 \caption[FIG8]{
 Same as Fig.~7, but for reflection. The insets in Fig.~(b) show
 the geometry of the pseudo-paths. Numbers next to insets give the number
 of revolutions inside the billiard.} 
 \end{figure}
The geometry of the microwave resonator used in the experiment \cite{blo02} is shown as
inset in Fig.~7.
The transmission and reflection amplitude were
taken in the frequency range $13\,$ GHz $ \le \nu \le 18\,$GHz, where only one
mode is propagating in the waveguides.
Figs.~7 and 8 show the experimental and calculated data for the
power spectra of the transmission and reflection amplitude, $|\tilde{T}_{11}(\ell)|^2$ and
$|\tilde{R}_{11}(\ell)|^2$. The important role of pseudo-paths in reproducing the correct peak pattern in good agreement with experimental results can be seen in both the transmission and the reflection spectra. Especially in  
$|\tilde{R}_{11}(\ell)|^2$ distinct pseudo-paths (see insets in Fig~8b) appear which were noted in Ref.~\cite{blo02}. For example, the peak at $\ell \approx 9$ (white arrow in Fig.~8a) is
caused by the trajectory with the length $\ell \approx 4.5$ which, after one revolution in the
billiard, is reflected back at the exit by the lead mouth, so that it continues for one more revolution. Thus its total length $\ell \approx 2 \times 4.5 = 9$. Of course, such non-classical
trajectories are not included in the standard semiclassical approximation
(Fig~8a). In this special case, the reflection at the open lead mouth is
specular or geometric. This peak would therefore also be present in the
semiclassical approximation suggested by Schwieters et al. \cite{schw96}. By
contrast, non-geometric reflections (``kinks'') by diffractive scattering are
also present, e.g.~the peak at $l=7.5$ (see inset in Fig.~8b). The latter
class is only contained in the PSCA.  
\section{Conclusions}
We have presented an extension of the pseudo-path semiclassical approximation
(PSCA) \cite{wir03} to billiards with arbitrary convex shape. A diagrammatic
expansion of the semiclassical Dyson equation for the Green's function in
terms of diffractive scattering diagrams is developed. The unperturbed Green's
function represents the sum over classical paths. Each encounter with the lead
(point contact) spawns new trajectories. By joining disjunct classical paths
due to non-geometric (diffractive) scattering a large number of trajectories
representing pseudo-paths results. The Dyson integral equation can be
converted to an algebraic matrix equation which can be solved by power series
expansion or inversion. Using the examples of a circular and a stadium
billiard we have shown numerically that the
path-length power spectra calculated by the PSCA overcome shortcomings of
standard semiclassical approximations by including an exponentially
proliferating number of pseudo-paths and converges towards quantum transport.
Moreover, we have presented a comparison between microwave billiard
experiments \cite{blo02}, the SCA and the PSCA calculation and find evidence
for contributions by pseudo-paths of PSCA to be present in the experimental
data.
We  expect that our
 pseudo-path semiclassical approximation will be able to address
 unresolved issues of semiclassical ballistic 
 transport such as the problem of weak localization 
 \cite{bar93a,stone94}, 
 the breakdown of symmetry of the autocorrelation function in 
 reflection and transmission, and the semiclassical description of quantum
 shot noise.\\[3ex]
  
\noindent {\bf ACKNOWLEDGEMENTS}\\
We thank H. Schanze, U. Kuhl and H.-J. St\"ockmann for providing the 
experimental data to allow 
comparison between the microcavity experiment and our pseudo-path 
semiclassical approximation.
Support by the grants FWF-SFB016 and FWF-P 17359 is gratefully acknowledged.
L. W. acknowledges support by the European Community
Network of Excellence NANOQUANTA (NMP4-CT-2004-500198).\\
\vspace{3ex}

\noindent {\bf APPENDIX: Fraunhofer approximation and its limitations}\\
The integrals along the transverse coordinate $(y)$ across the opening of the
point contact are evaluated in Fraunhofer diffraction approximation. The
dependence of the action in zero magnetic field, $k L_\alpha (y)$, on the
transverse coordinate $y$ is taken into account to first order Taylor series
expansion, 
\begin{equation}
\nonumber
L_{\alpha_i} (y) = L_{\alpha_i} + \sin (\theta_{\alpha_i}) y + \frac{\cos (\theta_{\alpha_i})}{L_{\alpha_i}}y^2 \cdots \, , \hspace{0.9cm} 
(A1)
\end{equation}
where $L_{\alpha} = L_\alpha (y = 0)$ is the length 
of the path that starts from the center of the lead mouth. Keeping all three
terms in (A1) leads to Fresnel diffraction integrals. Dropping the third term
results in a Fraunhofer diffraction approximation with the fundamental
integral 
\begin{equation}
\nonumber
 I^{FDA} (\theta, n)=\frac{1}{\sqrt{2 d}} \int \limits^{d/2}_{-d/2} e^{i(k_n+k \sin \theta) y} dy \, , 
\hspace{1.4cm}
(A2)
\end{equation}
  \begin{figure}[tb]
  \centering
  \includegraphics[draft=false,keepaspectratio=true,clip,%
                   width=0.95\linewidth]%
                   {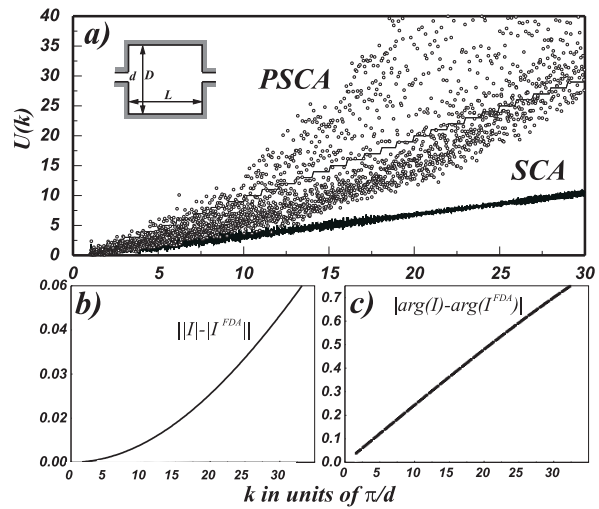}
 \caption[FIG9]{(a) Test of unitarity $U(k)$ on different levels of
   semiclassical approximations [SCA and PSCA ($K_{max}=10)]$ to the
   rectangular billiard ($D=L=\sqrt{4+\pi}$ and $d=$0.25). The staircase
   function shows the QM result where $U(k)= N(k)$ and $N(k)$ is the number of
   open modes. Fig.~(b) shows the discrepancy between the numerical
   integration of
   $I$ and the FDA corresponding to Eq.~(\ref{e:3.11}) (solid line) and
   Fig.~(c) the discrepancy in the phase angle.}
  \end{figure}
which, unlike Fresnel integrals, can be readily solved analytically in terms of elementary functions
\begin{equation}
\nonumber
 I^{FDA} (\theta, n)=\sqrt{\frac{2}{d}} \left( \frac{\sin [(k^i_n + k \sin \theta) d/2]}{k^i_n + k \sin \theta} \right) \, . \hspace{0.8cm}
 (A3)
\end{equation}
In terms of (A3) the reflection amplitude [Eq.~(\ref{e:3.12})] is given by 
\begin{equation}
\nonumber
r(\theta, \theta', k) = I^{FDA} (\theta, \theta',) k \sqrt{2 d (\cos \theta \cos\theta')} \, , 
\hspace{0.9cm}
(A4)
\end{equation}
with $\theta' = \sin^{-1} (k_n/k)$, the transmission amplitude
[Eq.~(\ref{e:4.5})] by
\begin{equation}
\nonumber
 t_m(\theta,k)= \sqrt{k \cos \theta } \left( I^{FDA} (\theta, m) + I^{FDA} (\theta, -m) \right) \, . 
 \hspace{0.1cm}(A5)
\end{equation}
The validity of the FDA hinges on the condition that the third term in
Eq.~(A1) is negligible. This term is of the order $(kd) (d/L_\alpha)
\lesssim kd \left( \frac{d}{D}\right)$ for short paths. Since in quantum
billiards the asymptotic far-field regime is never reached, the FDA, somewhat
counterintuitively, will fail for large $kd$. 
Indeed, in Fig. 9a we show a test of unitarity $U(k) = R(k) + T(k)$ on
different levels of semiclassical approximations to the rectangular billiard
(inset of Fig. 9a) where the breakdown of the FDA and consequently of the PSCA
for high modes $m$ ($m>10$) (or $kd \gtrsim$ 30) can clearly be seen. To
highlight the failure 
of the approximation used in Eq.~(A1) or (\ref{e:3.11}) we plotted in Figs. 9b
and 9c 
(solid lines) the discrepancy between the exact solution of the integral $I$ 
[LHS of Eq.~\eqref{e:3.11}] and the corresponding Fraunhofer approximation. 
For $L_{\alpha_1}=L_{\alpha_2}=L$ and 
$\theta_{\alpha_1}=\theta_{\alpha_2}=0$ as a function of $k$. Fig. 9b 
shows the difference in the absolute values $|I|-|I^{FDA}|$ and Fig. 9c the 
absolute discrepancy in the arguments $|\arg(I) - \arg(I^{FDA})|$. Since the
phase discrepancy reaches a fraction of unity for $m \gtrsim 16$ (see Fig.~9c)
random phases in the scattering amplitudes destroy path shadowing by
pseudo-paths and therefore cause the violation of unitarity (Fig.~9a). The
failure of the FDA has more dramatic consequences for the PSCA than for the
SCA. This is due to the fact that in PSCA the pseudo-path coupling leads to an
exponentially growing number of FDAs with an increasing number of diffractive
scatterings, in contrast to the linear scaling of the number of FDAs involved
within SCA. The point to be stressed is that this failure is not due to the
PSCA but due to the additional FDA. Applying more accurate diffraction
integrals in this regime is expected to remedy the problem.  


\begin{thebibliography}{99}
\bibitem{MRGM} The quantum mechanical calculations were performed using the modular recursive Green's function method (MRGM) \cite{rott00}
\end{thebibliography}
\end{document}